\documentclass[a4paper, aps, superscriptaddress, twocolumn, longbibliography]{revtex4-1}

\usepackage{amsfonts}
\usepackage{amssymb}
\usepackage{amsmath}    
\usepackage{graphicx}   
\usepackage{verbatim}   
\usepackage{color}     
\usepackage{subfigure} 
\usepackage{hyperref}  
\usepackage{braket}
\begin{document}

\title{Self-hybridization within non-Hermitian plasmonic systems}

\author{H. Louren\c co-Martins}
\thanks{These two authors contributed equally}
\affiliation{Laboratoire de Physique des Solides, Univ. Paris-Sud, CNRS UMR 8502, F-91405, Orsay, France}
\author{P. Das}
\thanks{These two authors contributed equally}
\affiliation{Laboratoire de Physique des Solides, Univ. Paris-Sud, CNRS UMR 8502, F-91405, Orsay, France}
\author{L. H. G. Tizei}
\affiliation{Laboratoire de Physique des Solides, Univ. Paris-Sud, CNRS UMR 8502, F-91405, Orsay, France}
\author{R. Weil}
\affiliation{Laboratoire de Physique des Solides, Univ. Paris-Sud, CNRS UMR 8502, F-91405, Orsay, France}
\author{M. Kociak}
\affiliation{Laboratoire de Physique des Solides, Univ. Paris-Sud, CNRS UMR 8502, F-91405, Orsay, France}

\date{\today}

\begin{abstract}
Common intuition in physics is based on the concept of orthogonal eigenmodes. Those are well defined solutions of Hermitian equations used to describe many physical situations, from quantum mechanics to acoustics. A large variety of non-Hermitian problems, including gravitational waves close to black holes or leaky electromagnetic cavities require the use of bi-orthogonal eigenbasis. Physical consequences of non-Hermiticity challenge our physical understanding \citep{Heiss2012,Brody2014,Lee2009}. However, the usual need to compensate for energy losses made the few successful attempts \citep{Stehmann,Lee2008,Doppler2016, Shin2016} to probe non-Hermiticity extremely complicated. We show that this issue can be overcome considering localized plasmonic systems. Indeed, since the non-Hermiticity in these systems does not stem from from temporal invariance breaking but from spatial symmetry breaking, its consequences can be observed more easily. We report on the theoretical and experimental evidence of non-Hermiticity induced strong coupling between surface plasmon modes of different orders within silver nano-crosses. The symmetry conditions for triggering this counter-intuitive self-hybridization phenomenon are provided. Similarly, observable effects  are expected to be observed in any system subtending bi-orthogonal eigenmodes.
\end{abstract}

\maketitle
Whatever be the field of physics described by an equation that has an Hermitian form - mechanics, acoustics, quantum mechanics, electromagnetism ... our intuition in linear physics is built upon the concept of eigenmodes.  Examples are endless: the vibrations of a guitar string are best understood as a superposition of the string eigenmodes and the properties of an atom can be simply deduced from its orbitals properties. It is thus tempting to adapt this concept to systems where their definition is harder to get, namely for non-Hermitian systems. \\
Indeed, many systems of importance are not Hermitian, but nevertheless can be advantageously described in terms of eigenmodes. A first class are open systems, for which energy is dissipated at infinity. Such systems span a wide range of physical situations, from gravity waves close to black holes to lasers cavities \citep{Leung1994,Ching1998,Leung1998}. Eigenmodes, generally called quasi-normal modes (QNMs), can be used to described them. QNMs are specially constructed so that time-invariance breaking does not prevent the constitution of a complete  basis. A second class are the plasmonic nanoparticles. In this case, the structure of the constituting equation is non-symmetric. Nevertheless a complete basis of eigenmodes can be deduced. For these two classes of problems, the price to pay to get an eigen-decomposition is that the basis is bi-orthogonal instead of being orthogonal. A full quantum theory of bi-orthogonal modes has been developed \citep{Brody2014}. Bi-orthogonality has few famous and exciting consequences, including the existence of special points known as exceptional points (EP) where both the energy and wavefunctions coalesce \citep{Heiss2012,Heiss2002a,Heiss2016,Seyranian2005}. EPs are usually associated with the apparition of non-trivial physical effects e.g. asymmetric mode switching \citep{Heiss2016}. Such effects have  only very recently been studied experimentally, in the case of open systems \citep{Stehmann,Doppler2016,Kodigala2016,Shin2016}.  However, these are extremely difficult to study because manipulating open systems eigenmodes requires exactly balancing dissipation \citep{Heiss2012, Zheng2010}.
Surprisingly, using localized surface plasmons (LSPs) to explore non-Hermiticity physics has not been reported, although balancing dissipation is not required in this case. Indeed, the need to use bi-orthogonal modes for describing LSPs physics has mostly been seen as an extra mathematical annoyance  \citep{Mayergoyz2005} that does not violate our common intuition. \\

\noindent Here, we show that the physics subtended by non-Hermiticity can be investigated theoretically and experimentally with LSPs. We explore the symmetry conditions required to evidence bi-orthogonality signatures in LSP systems. We show that both the surface plasmons equation's kernel symmetry and the overall system symmetry have to be tuned in that aim.  As a counter-intuitive consequence of non-Hermiticity, we predict the possibility of observing self-hybridization. This coupling within a nanoparticle concerns two bi-orthogonal modes of different orders, a situation impossible to occur with Hermitian systems. Studying silver nanocrosses through spatially resolved electron energy loss spectroscopy (EELS), we demonstrate that the effect is strong enough to be observed experimentally. Defining the relevant free-energy, we then draw an analogy between plasmons and other non-Hermitian systems such as open quantum cavities. Given the easily tunable parameters, we conclude that LSPs constitute an excellent platform for probing non-Hermitian physics.\\
\\
LSP resonances occur when a metallic nano-particle of arbitrary shape $S$ is excited by an external electric field. Within the quasi-static limit, Ouyang and Isaacson \citep{Ouyang1989} have shown that the plasmon modes are bi-orthogonal solutions of a Fredholm non-Hermitian eigenvalue problem which, in matrix form, reads:
\begin{align}
\label{eq:BEM}
\begin{split}
 F\; \ket{\sigma_m} &= \lambda_m \ket{\sigma_m} ,
\\
 \bra{\tau_m} \; F &= \lambda_m \bra{\tau_m} ,
\\
 F\left(\vec{r},\vec{r}\;'\right) &= -\dfrac{\vec{n}\left(\vec{r}\right).\left(\vec{r}-\vec{r}\;'\right)}{|\vec{r}-\vec{r}\;'|^3}\quad \textrm{where}\quad \vec{r},\vec{r}\; ' \in S
\end{split}
\end{align}

\noindent Where $\vec{n}\left(\vec{r}\;\right)$ is the outgoing normal at $\vec{r}$, $F$ is the normal derivative of the Coulomb Kernel, the right eigenvectors $\{\ket{\sigma_m}\}$ can be identified as surface charge densities, the left eigenvectors $\{\bra{\tau_m}\}$ are surface dipole densities projected along $\vec{n}$ and the eigenvalues $\{\lambda_m\}$ are dimensionless quantities associated to each pair of left-right eigenvectors. Hence, in contrast to systems recently considered \citep{Yin2013,Peng2014,Doppler2016,Shin2016,Hahn2016,Choi2017}, non-Hermiticity arises from the non-symmetry of $F$, which is still real. Solutions of equation \ref{eq:BEM} can be computed with the Boundary Element Method (BEM) \citep{GarciadeAbajo1997,Hohenester,Boudarham2012,Fredkin2003,Mayergoyz2005}. The integer $m$ indexes the modes by increasing values of $\lambda_m$. In the following, for the sake of simplicity, we will indifferently discuss the geometrical eigenvalues $\{\lambda_m\}$ or the plasmon eigenenergies $\{\omega_m\}$ assuming a one to one correspondence between the two spaces $\{\lambda_m\}\leftrightarrow\{\omega_m\}$. This is exact within the Drude model approximation for the dielectric function. It can be shown that $F$ is a quasi-Hermitian matrix \citep{Ouyang1989} (see supplementary section II). Therefore the $\{\lambda_m\}$ are real \citep{Ouyang1989}. Before going further, one must emphasize that there are fundamentally two types of symmetry involved in a plasmonic eigenproblem. The first one is \emph{the kernel symmetry} which controls the structure of the vector space solution and thus the (bi-)orthogonality of the plasmon modes. The second is \emph{the surface symmetry} (invariance of the surface charge or dipole distributions under any geometrical transformation) which may lead to additional properties of the plasmons. To avoid any confusion, in the following, we will refer to the kernel symmetry (the surface symmetry) as $F$-symmetry (respectively $S$-symmetry). As a practical example, in Fig. \ref{FIG:symmmetry}(a), we present $\sigma$ and $\tau$ (see methods and see supplementary information, section VI) corresponding to the two first eigenmodes of a $F$-symmetric surface (sphere) and a $F$-asymmetric surface (torus). As expected, for the sphere, the solutions are orthogonal and thus the left and right eigenvectors are identical while, in the case of the torus, the solutions are bi-orthogonal and the corresponding left and right eigenvectors are strikingly different. \\ 

\begin{figure}
    \includegraphics[width=\columnwidth]{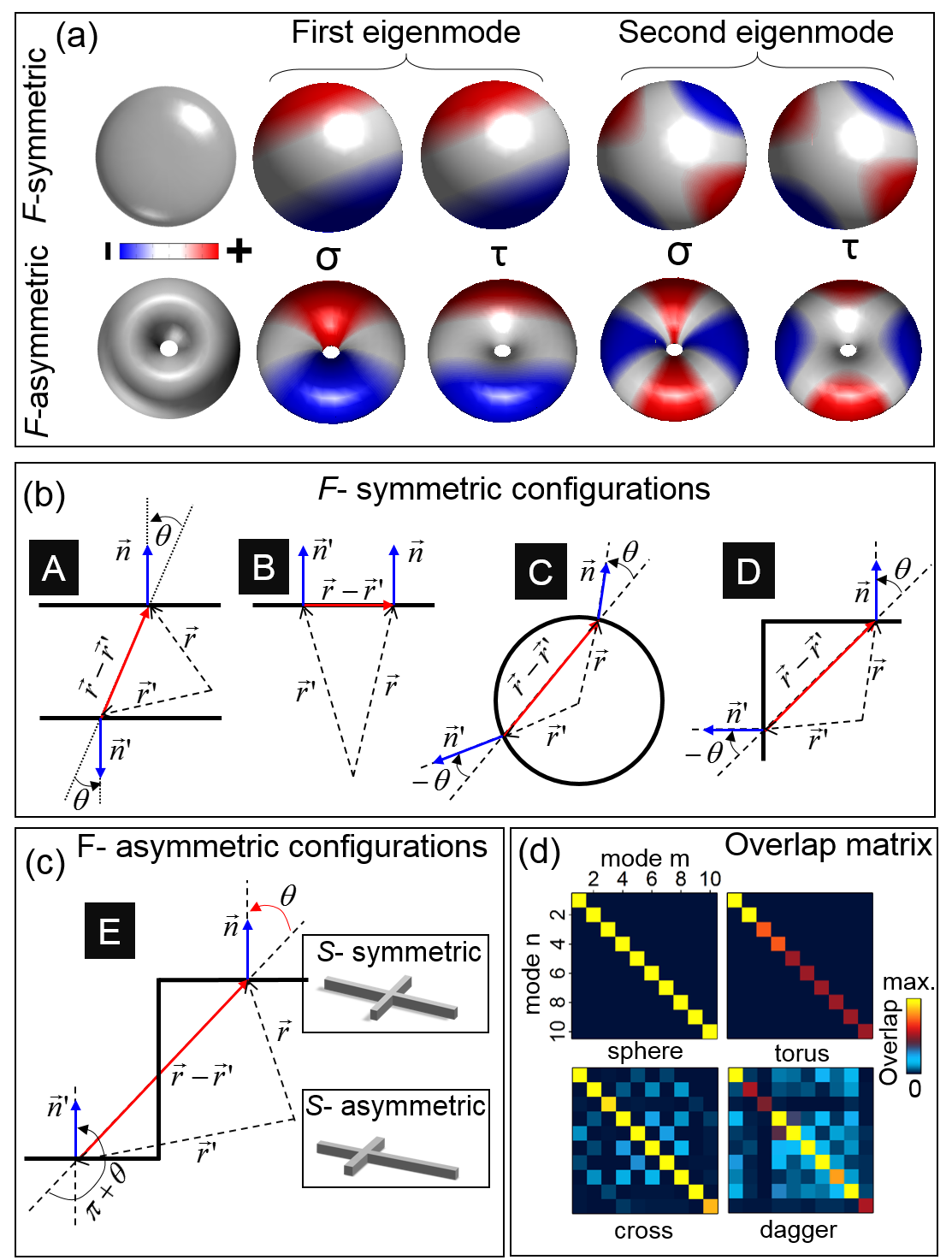}
    \caption{(a) Two first left and right eigenvectors of $F$-symmetric (sphere) and $F$-asymmetric (torus) surfaces. (b) Geometrical configurations of two normal vectors located on the surface leading to a symmetric contribution to the kernel. (c) Example of an asymmetric configuration corresponding e.g. to a cross or a dagger particle. (d) Overlap matrix between the 10 first eigenmodes of a sphere, a torus, a cross and a dagger.}
    \label{FIG:symmmetry}
\end{figure}

\noindent It can be shown that $F$-symmetric surface configurations ($F^\top=F$) satisfy (see supplementary section VI):
\begin{equation}
(\vec{n}\left(\vec{r}\;\right)+\vec{n}\left(\vec{r}\;'\right)).(\vec{r}-\vec{r}\;')=0
\end{equation}

\noindent where $\vec{n}$ (resp. $\vec{n}'$) is the normal vector in $\vec{r}$ (resp. $\vec{r}\;'$). In Fig. \ref{FIG:symmmetry}(b), we show four $F$-symmetric configurations A-D. From these, one can immediately deduce that a sphere (configuration C), a rod (configurations A, B and C), a cuboid \citep{Schmidt2016} (configurations A, B and D) or a disk (configurations C and D) are $F$-symmetric. Similarly, in Fig. \ref{FIG:symmmetry}(c), configuration E is obviously $F$-asymmetric and consequently the cross and the dagger (see inset) are $F$-asymmetric structures.\\ 
Moreover, when a surface is $F$-asymmetric, two right eigenvectors (or left) of different orders may have a non-zero spatial overlap, which may have dramatic consequences, as we demonstrate both theoretically and experimentally later. Therefore, quite counter-intuitively, two eigenmodes of the same nanoparticle and of different orders may interact while it is obviously impossible for orthogonal modes. In addition, we expect this interaction to be stronger as the overlap gets larger and, thus, one could formulate the following ansatz which will be justified later:
\begin{equation}
\Omega_{m,n}\propto T_{m,n}
\label{EQ:overlap}
\end{equation}
\noindent where $\Omega_{n,m}$ is the so-called classical Rabi energy of the two interacting modes $n$ and $m$. The overlap matrix $T_{m,n}=\braket{\tau_m| \tau_m} \braket{\sigma_m| \sigma_n}+\braket{\sigma_m| \sigma_m}\braket{\tau_m|\tau_n}$ (see supplementary section IV) thus constitutes a fundamental quantity to consider in the study of a bi-orthogonal systems. Let's emphasize that the hybridization mediated by the eigencharges we consider here is fundamentally different from the coupling in orthogonal systems mediated by the fields. On Fig. \ref{FIG:symmmetry}(d), we plotted the absolute value \footnote{Indeed, all the eigenvectors are determined up to an $e^{i\pi}$ phase. Consequently $T_{m,n}$ can indifferently take two values $\pm T_{m,n}$. To remove this uncertainty and to increase the dynamics of the colorscale, we plot the absolute value of the overlap matrix.} of the overlap matrix between the ten first eigenmodes of a sphere, a torus, a cross and a dagger. The sphere being $F$-symmetric, its overlap matrix obviously corresponds to the identity, as expected for orthogonal modes. As emphasized earlier, the torus is $F$-asymmetric but its matrix does not display any off-diagonal elements. This is a consequence of the strong $S$-symmetry (rotational invariance) of the torus shape which imposes $\braket{\sigma_m | \sigma_n}\propto\delta_{m,n}$ and $\braket{\tau_m | \tau_n}\propto\delta_{m,n}$. Consequently, although being $F$-asymmetric, the torus behaves essentially like an orthogonal system, the only difference being the absence of normalization of the elements on the diagonal. As discussed earlier, cross and dagger are two $F$-asymmetric structures which display weaker $S$-symmetry than the torus (see inset fig \ref{FIG:symmmetry}(c)). The cross is still centro-symmetric which imposes a null overlap between modes of different parity i.e. $\braket{\sigma_m | \sigma_m}=\braket{\tau_m | \tau_n}=0$ if $n+m$ is odd, resulting in the appearance of a checkerboard-like matrix. A comprehensive experimental and numerical study of the plasmonic cross system away from the hybridization point is developed in Ref. \citep{Das2017}. By shifting one arm of the cross, we break the centro-symmetry and the latter relation does not hold anymore. Consequently, the dagger overlap matrix has all its off-diagonal elements with non-null values (apart from accidental orthogonality, see supplementary section V). Therefore, $F$-asymmetry ensures that $\ket{\sigma}$ and $\ket{\tau}$ are different but does not guarantee that two different $\ket{\sigma}$ overlap. When the surface is $F$-asymmetric, the $S$-symmetry is the parameter controlling the overlap between modes of different orders.\\
\\
Bi-orthogonality enables eigenmodes of different orders to overlap thus to interact. This non-intuitive phenomenon of \emph{self-hybridization} should be accessible experimentally. More specifically, we expect the energy spectra of the LSPs to display an anti-crossing behavior as a function of a certain coupling parameter. In the present work (see methods), we used EELS in a scanning transmission electron microscope (STEM) as it has demonstrated its efficiency in probing plasmonic resonances with nanometric spatial resolution \citep{Nelayah2007}.\\

\begin{figure}
    \includegraphics[width=\columnwidth]{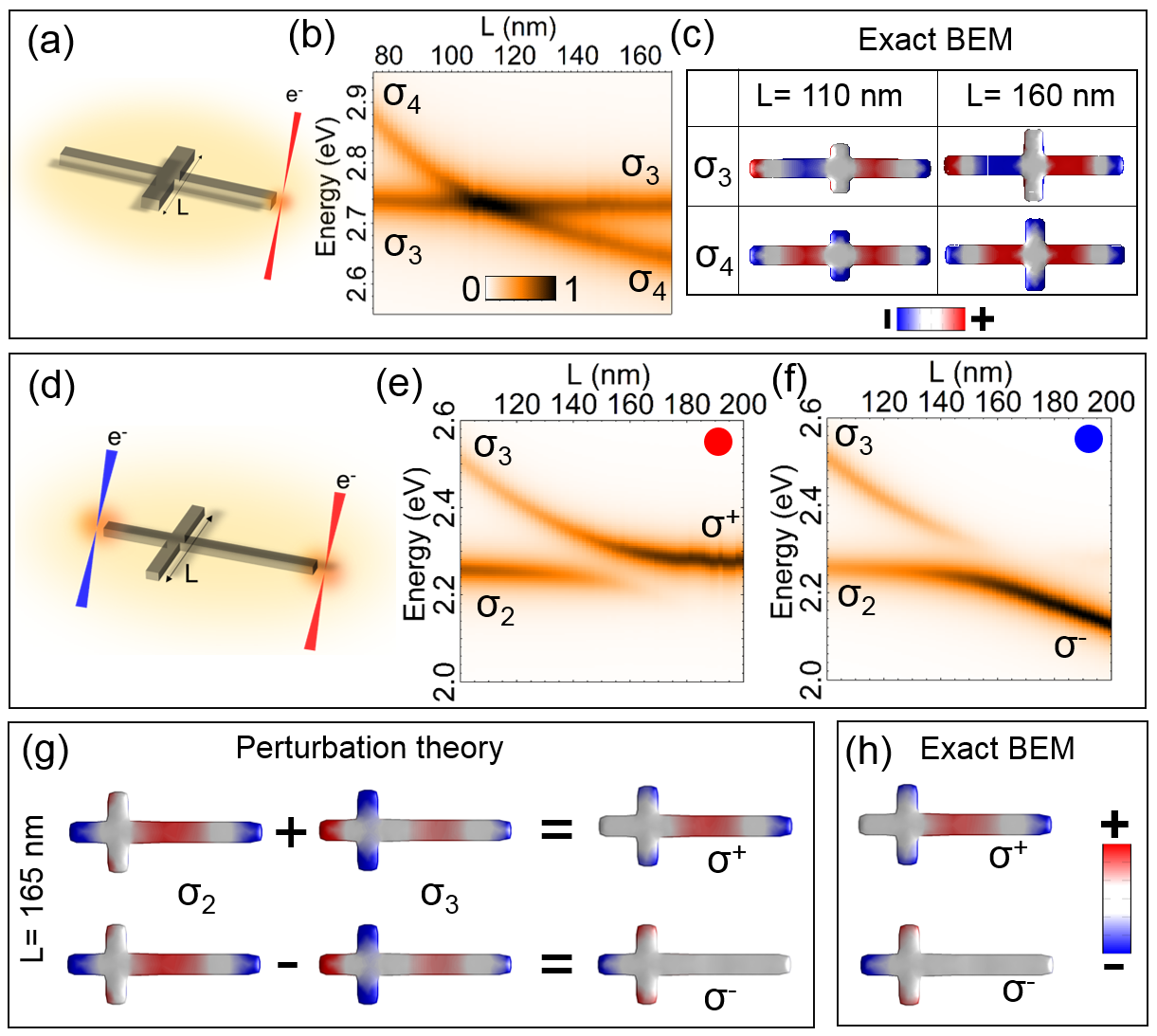}
    \caption{(a) Schematic representation of the EELS experiment on a cross. (b) Simulated EEL spectra taken at the position of the electron beam indicated in (a) as a function of $L$. (c) Simulated right eigenvectors corresponding to modes 3 and 4 at ($L$=110 nm) and after ($L$=160 nm) the crossing point. (d) Schematic representation of the EELS experiment on the dagger. (e) Simulated EELS spectra taken at the red position of the electron beam indicated in (d) as a function of $L$. (f) Simulated EEL spectra taken at the blue position of the electron beam indicated in (d) as a function of $L$. (g) Hybridized eigenvectors calculated at the anti-crossing point using first order perturbation theory. (h) Hybridized eigenvectors calculated at the anti-crossing point using the exact BEM. }
    \label{FIG:num}
\end{figure}

\noindent For the sake of the demonstration, we first consider 400 nm $\times$ $L$ silver cross with a 40$\times$40 nm square cross-section. The length $L$, which will be shown to be the relevant coupling parameter, is varied from 80 nm to 170 nm. The effect of the variation of $L$ on the eigenquantities can be modeled using first order perturbation theory. This approximation is formally derived for bi-orthogonal systems in \cite{Brody2014} and has been introduced for the boundary element method (BEM) by Tr\"ugler et al \citep{Trugler2011}. Within this approximation, one can treat a small geometrical deformation of the particle geometry as a perturbation of the kernel $F\rightarrow F+\delta F$, leading to a shift of the eigenvalues $\lambda_m^{(0)}\rightarrow  \lambda_m^{(0)}+\lambda_m^{(1)}$ but not to modification of the eigenvectors. This was elegantly used to analyze modes evolution when morphing a nano-triangle into a nano-disk \cite{Schmidt2016}. Within the perturbation theory, when two modes spectrally overlap, one has to take into account the possible hybridization between them \citep{Schmidt2014} by diagonalizing the typical Rabi-like matrix:  

\begin{equation}
M = 
\left(\begin{array}{cc} \lambda_m^{(0)}+\lambda_m^{(1)} & C_{m,n}\\
\\
C_{n,m} & \lambda_n^{(0)}+\lambda_n^{(1)} \end{array}\right)
\label{EQ:matrix}
\end{equation}

\noindent Using the convention of \citep{Zener1932,Brown2010,Novotny2010,Collins2014a}, we call \emph{diabatic} the eigenvectors of the unperturbed basis $\{\sigma_m^{(0)}, \tau_m^{(0)}\}$ in which $M$ is expressed in (\ref{EQ:matrix}), and \emph{adiabatic} the eigenvectors of the hybridized basis $\{\sigma_{m,n}^{\pm}, \tau_{m,n}^{\pm}\}$ in which $M$ is diagonal. At this point it is worth emphasizing, as it has been done in \citep{Schmidt2014}, that equation (\ref{EQ:matrix}) is similar to matrices encountered in linear combination of atomic orbital (LCAO) theory. This analogy is valid on a mathematical level but omits an important physical aspect of the problem. Indeed, LCAO theory describes the hybridization between orbitals belonging to different systems. Therefore, it can efficiently model dimer-like coupling where the two hybridized modes belong to two different and independent surfaces which can be either two monomers \citep{Nordlander2004} or two independent sub-surfaces of a large monomer \citep{Schmidt2016}. On the other hand, the self-hybridization process we describe here takes place within a single surface and would be comparable, for example, to the hybridization between $s$ and $p$ orbitals \emph{within} a single atom, and not between two atoms. Consequently, although mathematically analogous to LCAO, self-hybridization belongs to a specific universality class which is rather counter-intuitive. When the two modes are perfectly degenerated ($\lambda_m=\lambda_n$), one can show that the mixing term reads (see supplementary information, section IV) : 
\begin{equation}
C_{n,m}=\braket{\tau^{(0)}_n|\delta F|\sigma^{(0)}_m}=\lambda_m^{(1)}T_{n,m}
\end{equation}

\noindent Thus, one can immediately see that self-hybridization is only possible when $T_{n,m}\neq 0$ i.e when the system is bi-orthogonal. In other words, the surface defining the diabatic modes needs to be $F$-asymmetric while the $F$-symmetry of the  perturbative kernel $\delta F$ can be arbitrary.  The mixing term $C_{n,m}$ can be mapped in the energy space to the classical Rabi energy $\Omega_{n,m}$ which justifies the ansatz (\ref{EQ:overlap}). Contrary to $C_{n,m}$, $\Omega_{n,m}$ is an observable therefore, by measuring the energy splitting between two coupled modes, one can directly relate it to the degree of bi-orthogonality of a system.\\

\noindent In Fig. \ref{FIG:num}(a-b), using the exact BEM, we calculate the EELS spectra of the silver cross (when the beam impinges at one end of main axis, see Fig. \ref{FIG:num}(a)) as a function of the length $L$. When $L$ is small, the cross eigenmodes have the same spatial profile as the well known rod eigenmodes \citep{Collins2014a}. The corresponding eigenvectors $\ket{\sigma_n}$ thus display periodic profiles with $n$ nodes, see Fig. \ref{FIG:num}(c). When the length ($L$) of the arm is increased, the odd modes (odd $n$) which have no charge at the center remain almost unchanged while the even modes are expected to be red-shifted. Consequently, for particular values of $L$, modes of different parity can spectrally overlap, justifying the use of $L$ as a coupling parameter. As shown on Fig. \ref{FIG:num}(b), when $L$=110 nm, modes 3 and 4 spectrally overlap. However, although the cross is $F$-asymmetric, no sign of self-hybridization appears as the corresponding eigenvectors keep the same spatial profile at and after the crossing point (see Fig. \ref{FIG:num}(c)). As mentioned in Fig \ref{FIG:symmmetry}(d), this is due to the ($S$-)centro-symmetry of the cross which imposes a checkerboard form to the overlap matrix. In order to enable self-hybridization, one needs to break this accidental $S$-symmetry. To do so, we shift the position of the orthogonal growing arm to form a dagger-like geometry, cf Fig. \ref{FIG:num}(d). The position of the small arm of the dagger is chosen to correspond to a maximum of mode 3 and a node of mode 2. As for the cross, on Fig. \ref{FIG:num}(e-f), we calculate the EELS spectra as a function of $L$ at two different positions of the beam (the dagger being not left-right $S$-symmetric). The spectra display strong anti-crossing behavior, signature of the self-hybridization between modes 2 and 3. In order to validate our earlier interpretation, we calculate the adiabatic (hybridized) modes $\sigma_{\pm}$ both using the first order perturbation theory and the exact BEM on Fig. \ref{FIG:num}(g-h). The calculations are done exactly at the crossing point ($L$=165 nm) where the adiabatic modes are known to be equal mixtures of diabatic modes $\sigma_{\pm}=\sigma_{2}\pm\sigma_{3}$. The two results are in remarkably good agreement, proving that the perturbation theory gives a relevant picture of the self-hybridization physics.\\
One can highlight the exotic profile  of the hybrid modes. Particularly, the mode $\sigma_-$ strongly confines charges at one tip leading to a so-called hot-spot configuration which is of particular interest in a wide range of applications. Self-hybridization thus constitutes a very attractive procedure to design specific plasmonic states.\\
\\
Finally, we need to verify if the bi-orthogonality is a sufficiently strong phenomenon to be measured experimentally. To do so, we reproduced experimentally the simulations described in Fig. \ref{FIG:num}(d-f) by lithographing series of silver daggers with increasing $L$ and measuring the energy of mode 2 and 3 using the STEM-EELS technique (see methods). 

 \begin{figure}
 \includegraphics[width=\columnwidth]{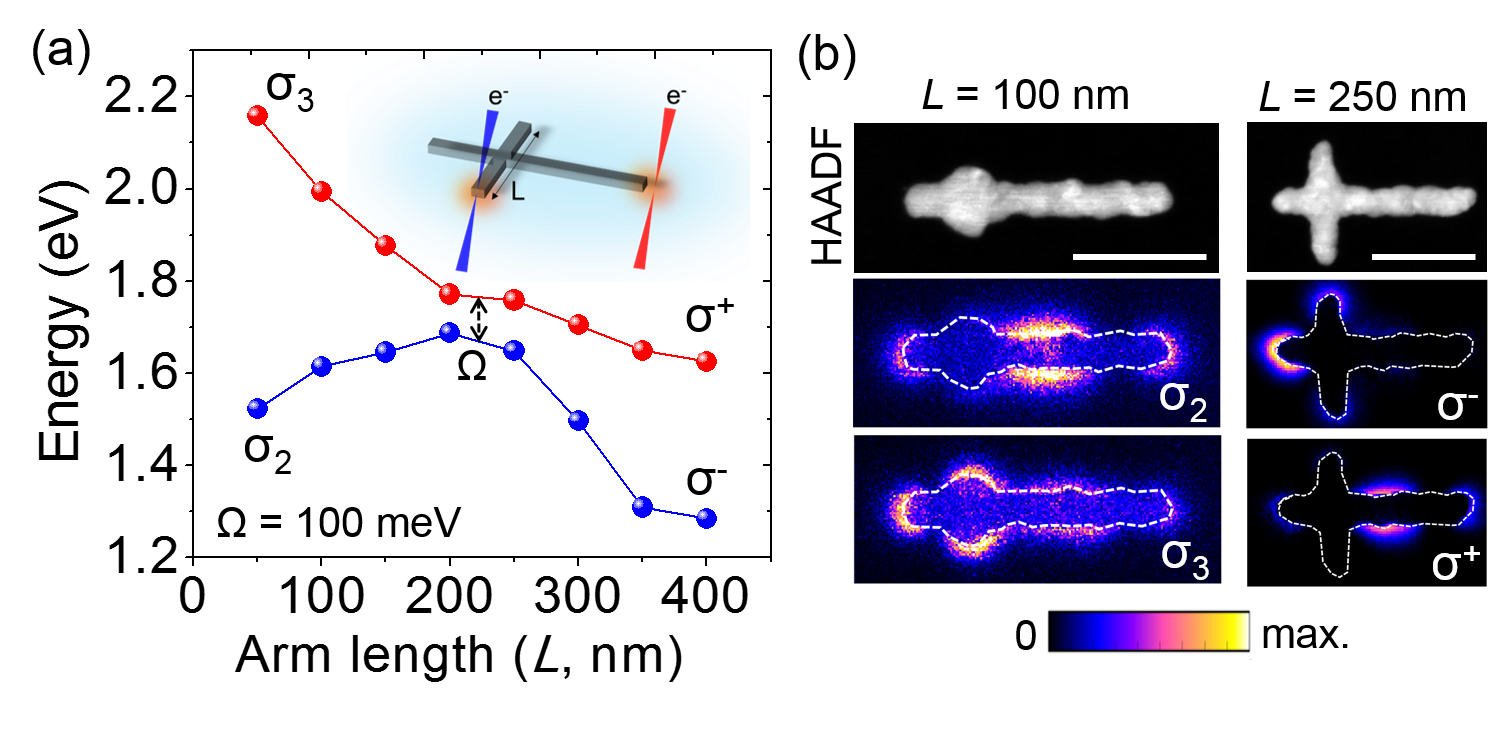}
     \caption{(a) Experimental plasmon energies of mode 2 and 3 (and their coupled $-$ and $+$ counterpart) measured using EELS for different arm lengths and taken at positions indicated on the schematic. (b) EELS filtered maps measured at energies corresponding to an uncoupled case  for $L=$100 nm and coupled case at $L=$165 nm. The top panel presents the high angle annular dark field (HAADF) image of the monomers. The scale bar corresponds to 200 nm.}
    \label{FIG:experiment}
\end{figure}

\begin{table*}[t]
\centering
\begin{tabular}{|c|c|c|} 
 \hline
  Physical quantities & Open quantum cavity & Plasmonics  \\
  \hline
  Time dependence & Dynamic & Static \\
  Kernel & Non-Hermitian Hamiltonian $H$ & Non-symmetric Coulomb kernel $F$ \\
  Eigenvalues & Complex energies $\omega_m$ & Real geometrical eigenvalues $\lambda_m$  \\
  Broken invariance & Time-reversal symmetry of $H$ & $F$ and $S$ spatial symmetry\\
  Constant characterizing the bi-orthogonality & Petermann factor K & Overlap matrix $T_{n,m}$ \\
  \hline
\end{tabular}
\caption{Table summing up the analogous quantities encountered in an open quantum system and a plasmonic system.}
\label{table}
\end{table*}

\noindent We report on Fig. \ref{FIG:experiment}(a) the energy of the two modes for different values of $L$. One can see that we reproduce the anti-crossing behavior calculated in Fig. \ref{FIG:num}. The lower branch ($\sigma_{-}$) and the upper branch ($\sigma_{+}$) of the hybridization scheme are separated by a coupling constant of $\Omega_{exp}\approx$100 meV which is a remarkably high value provided that the studied structures are lithographied polycrystalline particles. On Fig. \ref{FIG:experiment}(b) we report the EELS maps measured at the resonance energies for two different values of $L$. When $L$= 100 nm, the two modes display the spatial signature of the diabatic modes $\sigma_2$ and $\sigma_3$ showing that the two plasmons are not coupled. At $L$=250 nm, the coupling regime is clearly established as the two adiabatic plasmon modes display the characteristic spatial distributions expected from Fig. \ref{FIG:num}(h). The rigid redshift between the simulated curves Fig. \ref{FIG:num} and the experimental ones Fig. \ref{FIG:experiment} is due to retardation effects arising for large structures.\\

\noindent In conclusion, the self-hybridization is a strong and measurable phenomenon characteristic of the non-Hermiticity of the LSP's equation of motion. We should point out that this strong coupling regime has been reached by maximizing the overlap between the two eigenvectors. Therefore, the key parameter triggering the self-hybridization is the overlap matrix $T_{m,n}$. This quantity thus constitutes a measurement of the degree of bi-orthogonality of the system and therefore can be seen as classical analogue of the Petermann factor for lasers \citep{Savin2013}. Interestingly, we also note that S. Collins et al. \citep{Collins2014a} proposed phenomenologically that harmonic plasmonic modes within single nanorods could hybridize. The authors suggested that this could be the reason for an increase of intensity as certain nodes along the nanorod as measured by EELS. For symmetry reasons, nanorods eigenmodes cannot be degenerated, which may explain that S. Collins et al. could only measure very weak influence of self-hybridization. \\

\noindent In quantum mechanics, the appearance of non-Hermiticity is related to the broken time-reversal symmetry of the Hamiltonian. In complete analogy, one should identify which fundamental law controls the non-Hermiticity in classical plasmonic systems. For a given surface $S$, one can define a \emph{plasmonic energy functional} $\Xi$, which is the total surface charge-dipole interaction energy (see supplementary section I), as :   
\begin{equation}
\Xi=\dfrac{1}{4\pi}\oint_{S\times S} \dfrac{F^\top + F}{2}\; d\vec{s}\; d\vec{s}\;'
\end{equation}

\noindent The minimization of $\Xi$ should lead to appearance of new properties in the system. The surfaces which respect $\delta \Xi|_{S}=0$ are F-symmetric and the basis is orthogonal ($\{\sigma_n\}\propto\{\tau_n\}$). The surfaces which violate this minimization principle, $\delta \Xi|_{S}\neq0$, are non $F$-symmetric and thus the basis is bi-orthogonal ($\{\sigma_n\}\neq\{\tau_n\}$). While the time-reversal symmetry controls the Hermiticity of a Hamiltonian, the physical origin of the plasmonic bi-orthogonality is the violation of a variational principle.\\

\noindent In table \ref{table} we describe the analogy between plasmonic systems and open quantum cavity. This analogy can be extended to other types of QNM (see supplementary section II for comparison between LSPs and optical systems). Therefore we expect that strong manifestations of non-Hermiticity such as self-hybridization could be evidenced in other physical problems including gravitational waves, leaky electromagnetic cavity or acoustic cavities. On the other hand, we expect all the features of non-Hermitian systems to appear is LSP systems, particularly the presence of EPs.\\

\noindent \textit{Methods.} We produced a series of silver nano-crosses with an increasing arm length on 15 nm thin $Si_{3}N_{4}$ substrates. EELS measurements have been performed on a VG HB-501 STEM and a NION USTEM200 STEM. Both are equipped with a cold field emission gun (cFEG) operated at 60 or 100 kV and fitted with an homemade EELS detection system. Beam sizes were typically 0.7 and 0.15 nm and spectrometer entrance apertures were typically of the same angular-size as the incident beam. EEL spectrum images were deconvolved using a Richardson-Lucy algorithm \citep{Gloter2003} using typically 20 iterations resulting a zero-loss peak (ZLP) width with 0.15 eV. All the simulations have been carried out using the \texttt{MNPBEM} toolbox within the quasi-static formulation of the Maxwell equations and by considering only the 7th first eigenmodes of each structures.

\end{document}